\documentclass[showkeys,showpacs,prb,twocolumn, superscriptaddress,bibnotes]{revtex4}

\usepackage{amsmath}
\usepackage{color}
\usepackage{hyperref}
\usepackage{graphicx}
\usepackage{setspace}
\usepackage{tikz}

\renewcommand{\vec}[1]{{\boldsymbol{ #1}}}

\newcommand{\etal}{{\it et al.}}
\newcommand{\ie}{{i.e.}}
\newcommand{\Fig}{{Fig.}}

\begin{document}
\title{Current-induced switching of  magnetic 
tunnel junctions: Effects of field-like spin-transfer torque, pinned-layer magnetization orientation and temperature}

\author{R. K. Tiwari}
\affiliation{Institute of High Performance Computing, Agency for Science, Technology and Research, 1 Fusionopolis Way, \#16-16 Connexis, Singapore 138632}
\author{M. H. Jhon}
\affiliation{Institute of High Performance Computing, Agency for Science, Technology and Research, 1 Fusionopolis Way, \#16-16 Connexis, Singapore 138632}
\author{N. Ng}
\affiliation{Institute of High Performance Computing, Agency for Science, Technology and Research, 1 Fusionopolis Way, \#16-16 Connexis, Singapore 138632}
\author{D. J. Srolovitz}
\affiliation{Department of Materials Science and Engineering, Department of Mechanical Engineering and Applied Mechanics, University of Pennsylvania, Philadelphia, PA 19104, United States}
\author{Chee Kwan Gan}
\email{ganck@ihpc.a-star.edu.sg}
\affiliation{Institute of High Performance Computing, Agency for Science, Technology and Research, 1 Fusionopolis Way, \#16-16 Connexis, Singapore 138632}

\date{12 December 2013}

\begin{abstract}
We study current-induced switching in
magnetic tunnel junctions (MTJs) in the presence of a field-like
spin-transfer torque and titled pinned-layer magnetization
in the high current limit at finite temperature. 
We consider both the Slonczewski and field-like torques with coefficients $a_J$ and $b_J $, respectively.
At finite temperatures, $\sigma= b_J/a_J = \pm1$ leads to a smaller
mean switching time compared that with $\sigma = 0$.
The reduction of switching time in the
presence of the field-like term is due to the alignment effect (for $\sigma >
0$)  and the initial torque effect.  
\end{abstract}


\maketitle

The magnetic tunnel junction (MTJ) is the basic building block of magnetic
random access memory (MRAM) devices.  It consists of a thin, nonmagnetic
oxide film sandwiched between two magnetic layers. They exhibit tunnel
magnetoresistance, where the resistance of the junction depends on the
relative orientation of the magnetizations of two magnetic layers. An electric
current applies a spin-transfer torque (STT) that tends to align the
magnetizations of the magnetic layers.\cite{Slonczewski96v159,Berger96v54}
If the direction of the magnetization of one magnetic layer is pinned (fixed), a sufficiently strong STT
is able to switch the magnetization of the other (free) layer between two
states.\cite{Myers99v285} The spin-transfer torque $\tau_{\rm{STT}}$
due to a spin-polarized current has been phenomenologically described by
\cite{Heinonen10v105,Li08v100}
\begin{equation}
\label{eqn:stt}
\tau_{\rm STT}=a_J\vec{m}\times(\vec{m}\times\vec{m_p}) + b_J\vec{m}\times\vec{m_p}
\end{equation}
where the magnetizations of the free and pinned (fixed) layers
are denoted by unit vectors $\vec{m} $ and $\vec{m}_p$, respectively.
The first and second terms of the right-hand side of Eq.~\eqref{eqn:stt}
are called the Slonczewski and field-like terms, with prefactors $a_J$ and $b_J$, respectively. 
The subscript $J$ emphasizes the dependence of these prefactors on the current density $J$. 

The magnetization dynamics is the key factor in designing the switching 
of high-density scalable STT-MRAM. In particular, one requirement
for these devices is that they switch quickly between metastable
states. There have been several proposed design strategies
to engineer faster switching behavior, such as tilting the
magnetization of the pinned-layer away from the easy axis of the
free layer\cite{Zhu06v42,Zhou08v92,Zhou09v105,Zhou09v11} or adding
a second polarizer magnet.\cite{Kent04v84,Sbiaa09v105,Diao07v90}
However, these design strategies are typically evaluated assuming that
STT is controlled primarily by the Slonczewski-like term in Eq.~\eqref{eqn:stt}. Experiments have shown that the
spin-transfer torque in MTJs also can include a significant field-like
term,\cite{Petit07v98,Li08v100,Sankey08v04,Kubota08v04} unlike the case
of metallic spin valves.\cite{Xia02v65,Zimmler04v70} Although $b_J$  can
be $10-100$\% of $a_J$,\cite{Theodonis06v97,Slonczewski07v310,Kubota08v04,Li08v100,Petit07v98,Sankey08v04}
the magnitude and sign of $b_J$ is not as well understood
as that of $a_J$~\cite{Li08v100}  and the exact bias dependence
of $b_J$ is not clear.  For example, Sankey \etal~\cite{Sankey08v04}
and Kubota \etal~\cite{Kubota08v04} found $b_J$
to be quadratic in the bias, while Petit \etal~\cite{Petit07v98} found it to be linear.

Zhou\cite{Zhou11v109} obtained theoretical
limits for the switching current density and switching time in the presence
of the $b_J$ term.  Recently Butler \etal\cite{Butler12v48} applied the Fokker-Planck approach
to the switching distributions of spin-torque devices to find the long-time nonswitching (switching)
probability for the write (read) process. To do this, they considered
the effect of a dimensionless current that lumps together various
effects including $a_J$ and $b_J$.  In this work, we use a different
approach to study the current-induced switching behavior of a MTJ in the
presence of the field-like term with macromagnetic simulations based
on a stochastic Landau-Lifshitz-Gilbert equation.  In the absence of a
field-like term, previous studies have demonstrated the importance
of tilting the pinned-layer\cite{Zhu06v42,Zhou08v92,Zhou09v105} and
temperature\cite{Diao07v19,Braganca05v87} on the switching dynamics;
here we consider both of these effects. We also study the switching behavior 
for the case that $b_J$ is assumed to vary quadratically with $J$.

\begin{figure}
\includegraphics[clip,width=7.2cm]{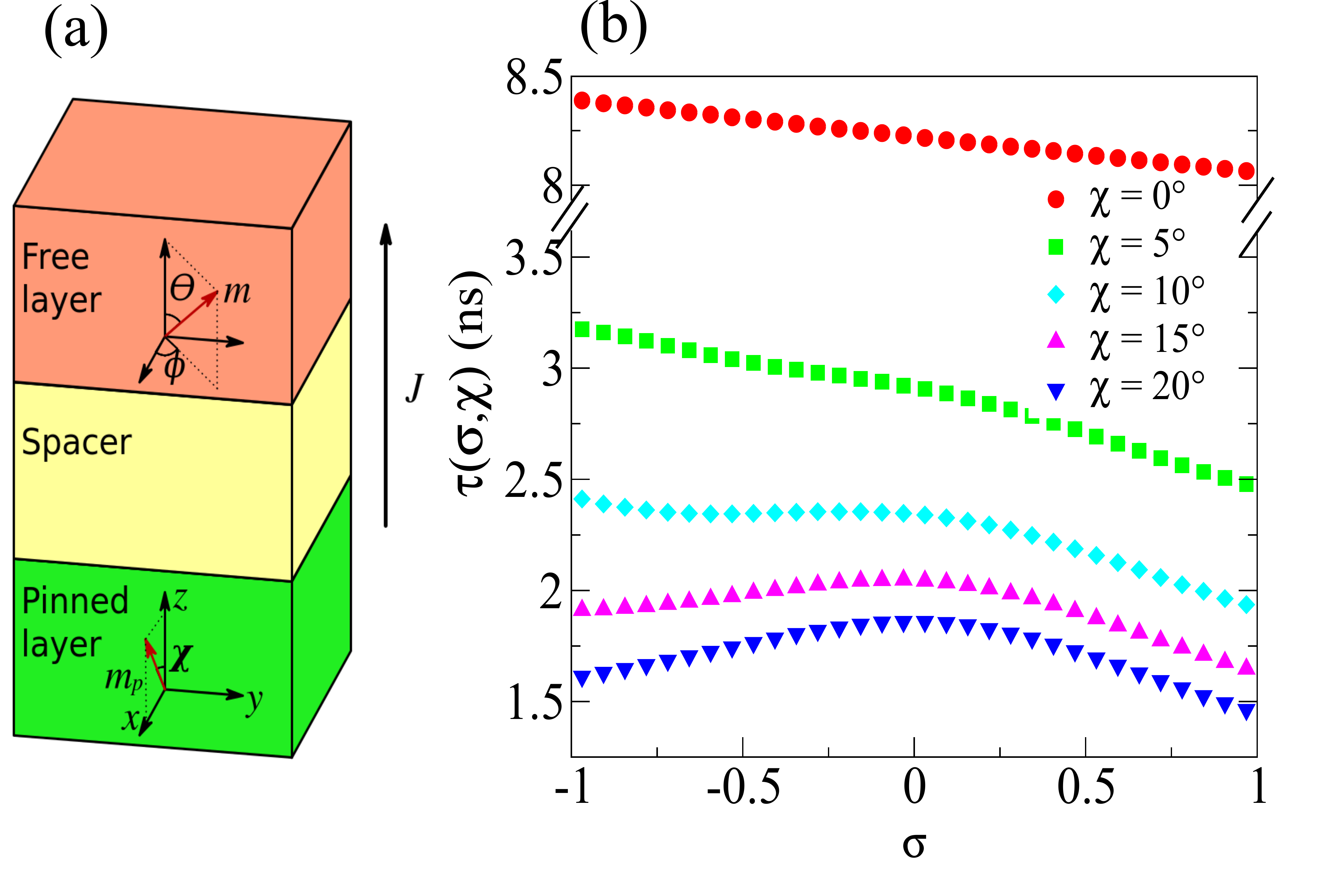}
\caption{(a) Schematic illustration of the device studied here.
The red arrows show the orientation of the magnetization vectors of the 
free and pinned-layers. The pinned-layer magnetization 
is constrained to lie in the $x$-$z$ plane making 
an angle $\chi$ with the $z$-axis. (b)  Switching
time, $\tau({\sigma,\chi})$ as 
a function of $\sigma = b_J/a_J$ for different values of $\chi$, the
tilt angle of the pinned-layer.}
\label{fig:figure1}
\end{figure}

We consider the free layer of the
MTJ as schematically illustrated in Fig.~\ref{fig:figure1}a.  The red
arrows illustrate $\vec{m}$ and $\vec{m}_p$, unit vectors parallel
to the magnetization of the free and pinned-layers, respectively.
The orientation of $\vec{m}$ is described by the polar angle $\theta$
and the azimuthal angle $\phi$.  $\vec{m}_p$ is constrained (with no loss of generality) to the
$xz$ plane,  makes an angle $\chi$ with the $z$ axis. We simulate the
dynamics of the magnetization of the free layer by integrating the
Landau-Lifshitz-Gilbert equation,
\begin{equation}  
\frac{d\vec{m}}{dt} = -\gamma \vec{m} \times \vec{H}_{\rm eff}  + \alpha \vec{m} \times \frac{d\vec{m}}{dt} - \tau_{\rm{STT}} 
\end{equation}
where $\gamma$ is the gyromagnetic ratio, $\alpha$ is the Gilbert
damping constant, and $\vec{H}_{\rm eff}$
is the effective field.
The spin-transfer torque $\tau_{\rm{STT}}$ is described by Eq.~\eqref{eqn:stt}.
We use a convention where $J > 0$ corresponds to 
electrons moving in the positive direction\cite{Xiao07v102}
(\ie, the conventional current density has a negative sign when $J > 0$).
For the Slonczewski term, we use the standard expression
\begin{equation}
a_J = \frac{\gamma\hbar J \epsilon}{\mu_0 e d M_s}
\end{equation}
\noindent
where $e$, $d$, and $M_s$ represent the elementary charge, thickness of the
free layer, and saturation magnetization, respectively.  $\epsilon$
characterizes the angular dependence of the Slonczewski term where
\begin{equation} \epsilon = \frac{P \Lambda^2}{(\Lambda^2 + 1) +
(\Lambda^2 -1)(\vec{m} \cdot\vec{m}_p)} \end{equation} \noindent where
$P$ and $\Lambda$ are dimensionless quantities that determine the
spin polarization efficiency.\cite{Xiao04v70} We introduce $\sigma =
b_J /a_J $ to characterize the relative strength of $b_J$ relative to
$a_J$. 
We restrict the value of $\sigma$ from $-1$ to $1$ for all $J$ investigated
since experimentally $b_J$ has been found to be $10-100\%$ of $a_J $.\cite{Kubota08v04,Li08v100,Petit07v98,Sankey08v04}

For the finite-temperature studies, we augment $\vec{H}_{\rm{eff}}$ with a random
fluctuating field $\vec{H}_r$ whose 
statistical properties are given by \cite{Garcia98v58}
$\langle H_r^i(t) \rangle = 0  $
and 
\begin{equation}
\langle H_r^i(t) H_r^j(t') \rangle = \frac{2\alpha k_B T}{(1+\alpha^2) \gamma' M_s \mu_0 V} \delta_{ij} \delta(t-t'),
\end{equation}

\noindent
where $i$ and $j$ are Cartesian indices. $\gamma' =
\gamma/(1+\alpha^2)$, $V$ is the volume of the cell, and $T$ is the
absolute temperature.  $\langle \cdot \cdot \cdot \rangle$ denotes the
time average of the enclosed quantity.

For all calculations, we set the thickness of the free, spacer, and 
pinned-layers to $1.3$~nm and the cross-sectional area to 
$100\times 100$~nm$^2$.  The saturation magnetizations for the
free and pinned-layers are $M_s = 1200$~kA/m, the uniaxial anisotropy
constant $K_u=2.83$~kJ/m$^3$, the Gilbert damping constant $\alpha=0.02$,
and the spin torque parameters are $P=0.40$ and $\Lambda=1$.   $M_s$ 
is consistent with previous reported values 
for a similar device \cite{Ikeda10v09}.  Unless otherwise stated, we set $J=-7.64
\times 10^{10}$~A/m$^2$ to fall within the high current limit to ensure
switching times of less than $ 10$~ns.  The switching time
$\tau$ is the time for $m_z$ to first change sign.

To understand the switching dynamics, 
we first perform deterministic $0$~K calculations, varying both $\sigma
$ and the pinned-layer tilt angle $\chi$.  We note that an average value
of $\theta$ of $0.01^\circ$ corresponds to $1\times 10^{-4}$~K using our
device parameters after thermal equilibration.  Therefore, we set the initial conditions to
$\theta =0.01^\circ$ and $\phi = 0^\circ$.  \Fig~\ref{fig:figure1}b
shows that $\tau(\sigma,\chi)$ generally decreases with increasing $\chi$
and $|\sigma|$ except for small $\chi$ and negative $\sigma$. This can
be understood by noting that the field-like term acts as an external
field $\vec{H}_{\rm{FL}}=b_J\vec{m}_p/\gamma$, oriented parallel
(anti-parallel) to the pinned-layer magnetization for positive (negative)
$b_J$.  The effect of $\vec{H}_{\rm{FL}}$ is to create a torque that
is proportional to $\sin \theta'$, where $\theta'$ is the angle between
$\vec{m}$ and $\vec{m}_p$, that may move $\vec{m}$ out of the easy axis
and thus assist in switching.  We call this the initial torque effect.
$\vec{H}_{\rm{FL}}$ also tends to align $\vec{m}$ in the direction
of $\vec{H}_{\rm{FL}}$ which may also further assist in switching. We
call this the alignment effect. Specifically, we first consider parallel (P) to
anti-parallel (AP) switching, which requires $a_J<0$ (see \Fig~\ref{fig:figure1}a).
If $\sigma>0$ then $b_J<0$ and $\vec{H}_{\rm{FL}}$ is anti-parallel
to the pinned-layer magnetization that helps in switching. Thus,
both the initial torque and the alignment effect aids P to AP
switching. With increasing $\chi$, the initial torque increases and
$\tau(\sigma,\chi)$ decreases with $\chi$.  Similarly, with increasing
$\sigma$ both the initial torque and alignment effect increases, leading
to a decrease in $\tau(\sigma,\chi)$ with $\sigma$.

Next, we consider the $\sigma<0$ case in P to AP switching, where $b_J>0$ 
and $\vec{H}_{\rm{FL}}$ is parallel to the fixed layer magnetization. Here, 
the alignment effect opposes switching. For small $\chi$ the initial
torque is small and we see an increase in $\tau(\sigma,\chi)$. For large $\chi$,
the initial torque may overcome the alignment effect and this may lead to 
an overall decrease in $\tau(\sigma,\chi)$. The result of $\tau(\sigma,\chi)$ 
for AP to P switching is exactly the same as that for P to AP since the 
arguments are the same for both cases. Our analysis suggests that for good 
P to AP or AP to P  switching performance, the sign of $b_J$ should
change in such a way that $\sigma$ is always positive. 
 
We next consider the effect of temperature on switching dynamics
using two schemes.  In the first, we simulate the system at $0$~K
so that the trajectories are fully deterministic, but with the initial
configurations taken as a result of thermalization at $300$~K. This is
to provide a better understanding for the second scheme where the 
simulation is performed at $300$~K (non-deterministic trajectories)
and the initial configurations are the result of $300$~K thermalization.  
Statistics are collected from $512$ identically prepared systems each equilibrated
for $60$~ns.

\begin{figure}
\includegraphics[clip,width=8.0cm]{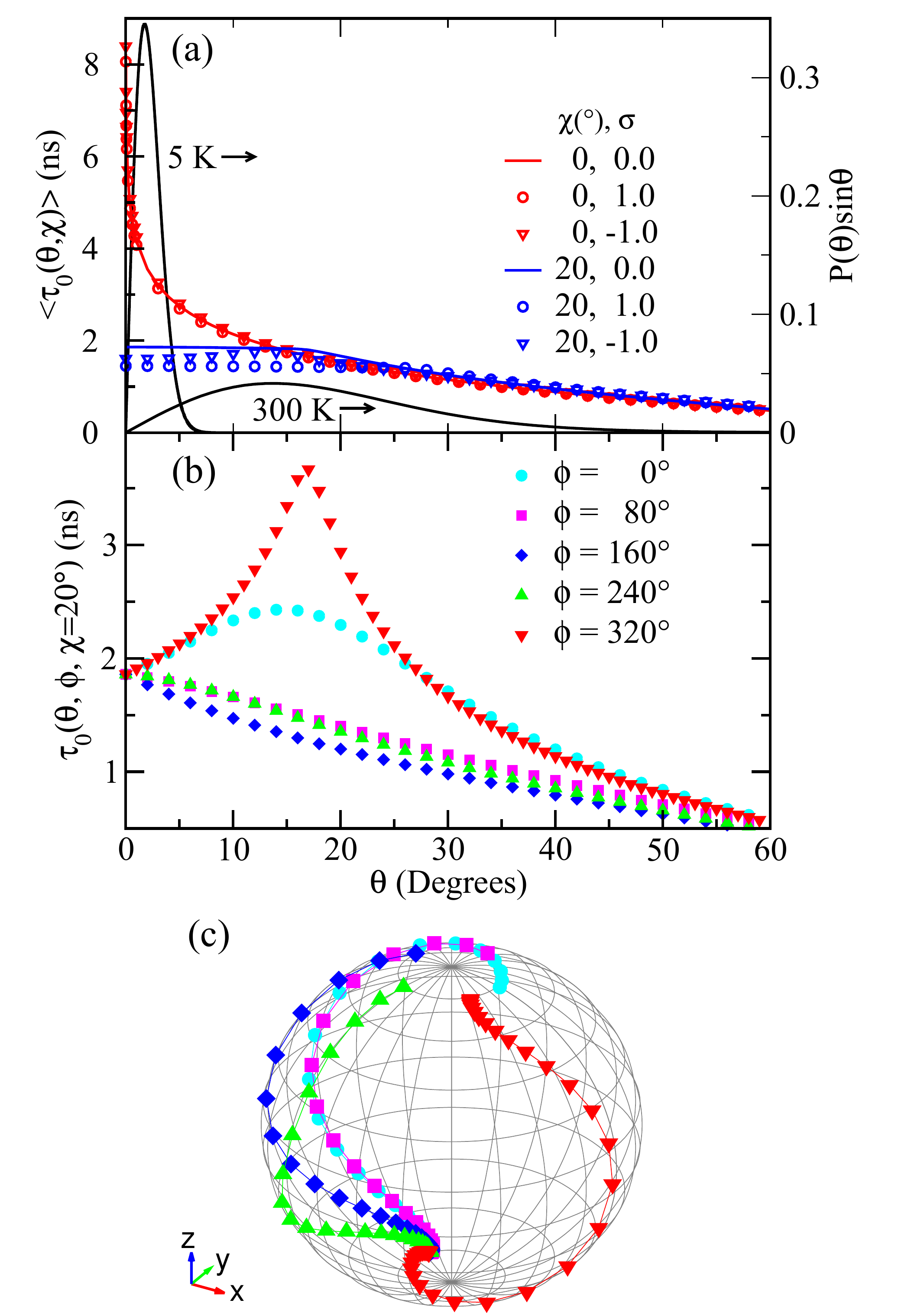}
\caption{
(a) $\langle \tau(\theta,\chi) \rangle_\phi$ versus $\theta$ for $\chi=0^\circ$ 
and $\chi=20^\circ$ and $\sigma=0,\pm1$. $P(\theta) \sin \theta$ at $T= 300$~K
and $5$~K are shown as dashed, black lines.
(b) $\tau(\theta,\phi,\chi=20^\circ)$ versus $\theta$ for several values of $\phi$.
(c) The trajectories of  $\vec m$ with $\sigma=0$ and $\theta=17^\circ$
for several $\phi$ values. The dots on the trajectories are drawn at equal time 
intervals. Densely populated dots at the beginning of the $\phi=320^\circ$ trajectory 
are due to the small torque in this configuration.
}
\label{fig:figure2}
\end{figure}

In the first scheme, the mean switching time, $\langle \tau_0(\chi)
\rangle$ (where the subscript indicates that the switching time is
calculated at $0$~K) is given by
\begin{equation}
\label{eqn:meantau}
\langle \tau_0 (\chi) \rangle  = \frac{1}{2\pi} \int_{0}^{\pi/2} d\theta\  P(\theta) \sin \theta  \int_{0}^{2\pi} d\phi\  \tau_0(\theta, \phi, \chi) ,
\end{equation}
where $\tau_0(\theta,\phi,\chi)$ denotes the
deterministic switching time for an initial spin configuration
$(\theta,\phi)$ and a tilt angle $\chi$ of the pinned-layer.  The
equilibrium probability distribution of the magnetization vector $\vec
m$ making an angle $\theta$ with the $z$-axis (averaged over $\phi$)
$P(\theta)\sin \theta$ (shown in \Fig~\ref{fig:figure2}a),
accounts for the initial spin configuration at $T=300$~K
and $T=5$~K.  On introducing $\langle \tau_0 (\theta, \chi) \rangle_\phi =
(2\pi)^{-1}\int_0^{2\pi}d\phi\  \tau_0(\theta,\phi,\chi)$ as the switching
time averaged over $\phi$, we reexpress Eq.~\eqref{eqn:meantau} as
\begin{equation}
\label{eqn:meantauphiaveraged}
\langle \tau_0 (\chi) \rangle = \int_{0}^{\pi/2} d\theta\   P(\theta) \sin \theta \langle \tau_0 (\theta, \chi) \rangle_\phi .
\end{equation}

Written in this form, the effect of the temperature comes in only
through $P(\theta) \sin \theta$, while allowing us to study 
$\langle \tau_0 (\theta,\chi)\rangle_\phi$ which is shown in
\Fig~\ref{fig:figure2}a for $\chi=0^\circ$ and $\chi=20^\circ$ with
$\sigma=0,\pm1$. First, we consider the case when $\sigma=0$ (\ie, no
field-like term). $\langle \tau_0 (\theta, \chi=0^\circ) \rangle_\phi$
is  monotonically decreasing function of $\theta$ because initial
torque scales as $\sin \theta$.  $\langle \tau_0 (\theta, \chi=20^\circ)
\rangle_\phi$, however, remains constant for $\theta$ from $0$ to $\sim
20^\circ$.  We note that when $\theta$ varies from $0$ to $\pi/2$,
$\vec m$ sweeps out progressively larger circles when $\phi$ changes
from $0$ to $2\pi$. As shown in \Fig~\ref{fig:figure2}b, as  $\theta$
approaches $\chi=20^\circ$, some configurations $\vec m$ may almost
align with $\vec{m}_p$ that results in a small torque that drastically
increases the switching time.  However, for other configurations where
$\vec{m}$ points away from $\vec{m}_p$, the torque increases and the
switching time decreases.  The overall effect is that the average $\langle
\tau_0(\theta, \chi) \rangle_\phi$ is nearly constant from $0$ to $\chi$
(see \Fig~\ref{fig:figure2}a for the case of $\chi=20^\circ$). The
trajectories of $\vec m$ for $\theta=17^\circ$ and $\chi=20^\circ$
for several $\phi$ values are shown in \Fig~\ref{fig:figure2}c which
explains a large variation in $\tau_0(\theta,\phi,\chi)$ shown in
\Fig~\ref{fig:figure2}b for $\theta \approx \chi$.

The mean switching times obtained using Eq.~\eqref{eqn:meantauphiaveraged}
for $\chi=0^\circ$ and $\chi=20^\circ$ is $1.75$~ns  and $1.63$~ns,
respectively. Thus we see that there is a slight decrease in the
mean switching time when $\chi$ varies from $0^\circ$ to $20^\circ$.
This difference is due to a smaller $\langle \tau_0 (\theta,\chi)
\rangle_\phi$ for $\chi=20^\circ$ compared to that for $\chi=0^\circ$
(see \Fig~\ref{fig:figure2}a). 
Note that the effect of large $\langle \tau_0 (\theta,\chi) \rangle_\phi$ difference between
$\chi = 0^\circ$ and $20^\circ$ when $\theta$ is small ($ <\sim 5^\circ$) as shown
in \Fig~\ref{fig:figure2}a  in determining the mean switching time $\langle \tau_0 (\chi) \rangle$
is somewhat suppressed by the small $P(\theta)\sin\theta$ values for these $\theta$ at $T=300$~K.

We now consider the effect of the field-like term in determining
the mean switching time.  For $\chi=20^\circ$, $\sigma=1\ (\sigma=-1)$
leads to a mean switching time of $1.37$~ns ($1.51$~ns); this is
smaller than the $1.63$~ns time at $\sigma=0$. When $\chi=0^\circ$
there is no reduction in the mean switching time in the presence of
the field-like term because $\langle \tau_0 (\theta,
\chi=0^\circ) \rangle_\phi$ is largely unaffected by
the presence of the field-like term (see \Fig~\ref{fig:figure2}a)
since it does not produce a torque in the $z$-direction
when $\chi=0^\circ$ for all values of $\theta$ and $\phi$.  We note
that $\langle \tau_0(\theta<0.1^\circ,\chi) \rangle_\phi$ (see
\Fig~\ref{fig:figure2}a for $\chi=0^\circ$ and $20^\circ$,
and $\sigma = 0,\pm1$) are consistent with the results shown in
\Fig~\ref{fig:figure1}b.

\begin{figure}
\includegraphics[width=7.0cm]{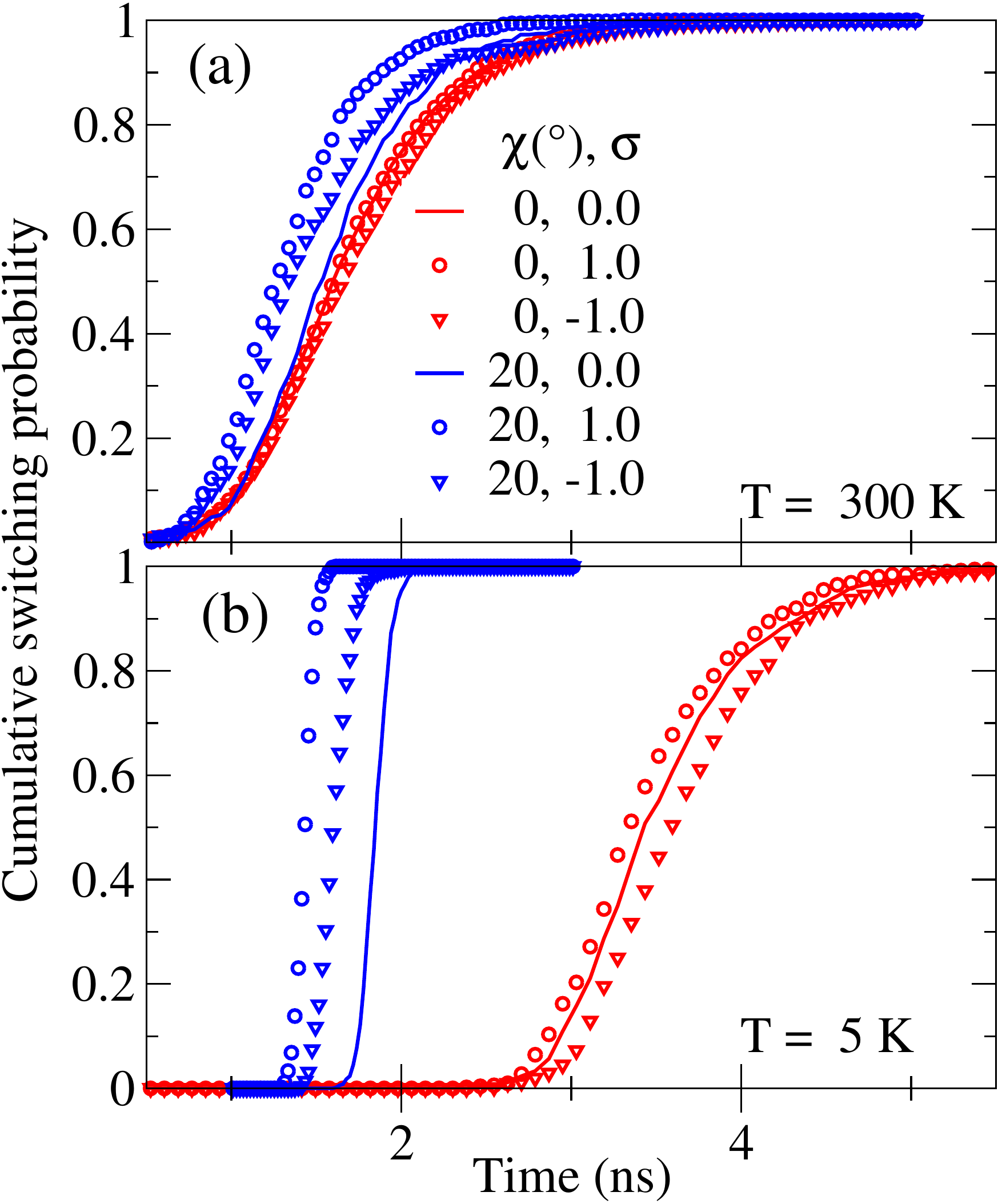}
\caption{\label{fig:figure3}Cumulative switching probability distributions
for $\chi = 0^\circ$ and $20^\circ$
with $\sigma = 0$ and $\pm1$ at (a) $300$~K and (b) $5$~K.}
\end{figure}

The switching statistics obtained using the second scheme are shown in
\Fig~\ref{fig:figure3}a for $T=300$~K. We find that in the absence
of the field-like term, the mean switching times (as deduced from the
cumulative distribution probability curve) are $1.71$~ns and $1.61$~ns for
$\chi=0^\circ$ and $\chi=20^\circ$, respectively.  These are in
reasonable agreement with the respective values of $1.75$~ns and $1.63$~ns
obtained using the first scheme. In the presence of the field-like term, the
mean switching time remains nearly unchanged compared to the case when
the field-like term is absent for $\chi=0^\circ$. For $\chi=20^\circ$,
the presence of the field-like term with $\sigma=1\ (\sigma=-1)$ results
in  mean switching times of $1.35$~ns ($1.50$~ns), which is also
consistent with the results obtained in the first scheme. We conclude
that temperature changes the switching time mainly through its effect
on the initial spin configuration. 
Since the effect of initial torque is
large for large $\chi$, we expect that the mean switching time
will decrease when $\chi$ is increased. Indeed,
with a typical value of $\chi = 30^\circ$ we find that the mean
switching times obtained with the second scheme for $\sigma=0, 1, -1$
are $1.51$, $1.20$, and $1.29$~ns, respectively. These values 
are consistently less than the corresponding values 
for $\chi = 20^\circ $ or $\chi = 0^\circ$.

Similar analysis can also be performed at $T=5$~K where $P(\theta)\sin
\theta$ peaks at small angles ($\sim 2^\circ$). With $ \langle
\tau_0(\theta,\chi) \rangle_{\phi}$ (see \Fig~\ref{fig:figure2}a), we
anticipate that $\chi=20^\circ$ delivers a much lower mean switching time
compared to $\chi=0^\circ$. Indeed, \Fig~\ref{fig:figure3}b shows the
mean switching time can be dramatically reduced by tilting the $\chi$ from
$0^\circ$ to $20^\circ$. Also at $\chi = 20^\circ$, $\sigma = \pm 1$ leads
to smaller switching times compared to $\sigma=0$
since the field-like term introduces a large initial torque.

\begin{figure}
\vspace{10pt}
\includegraphics[clip,width=7.2cm]{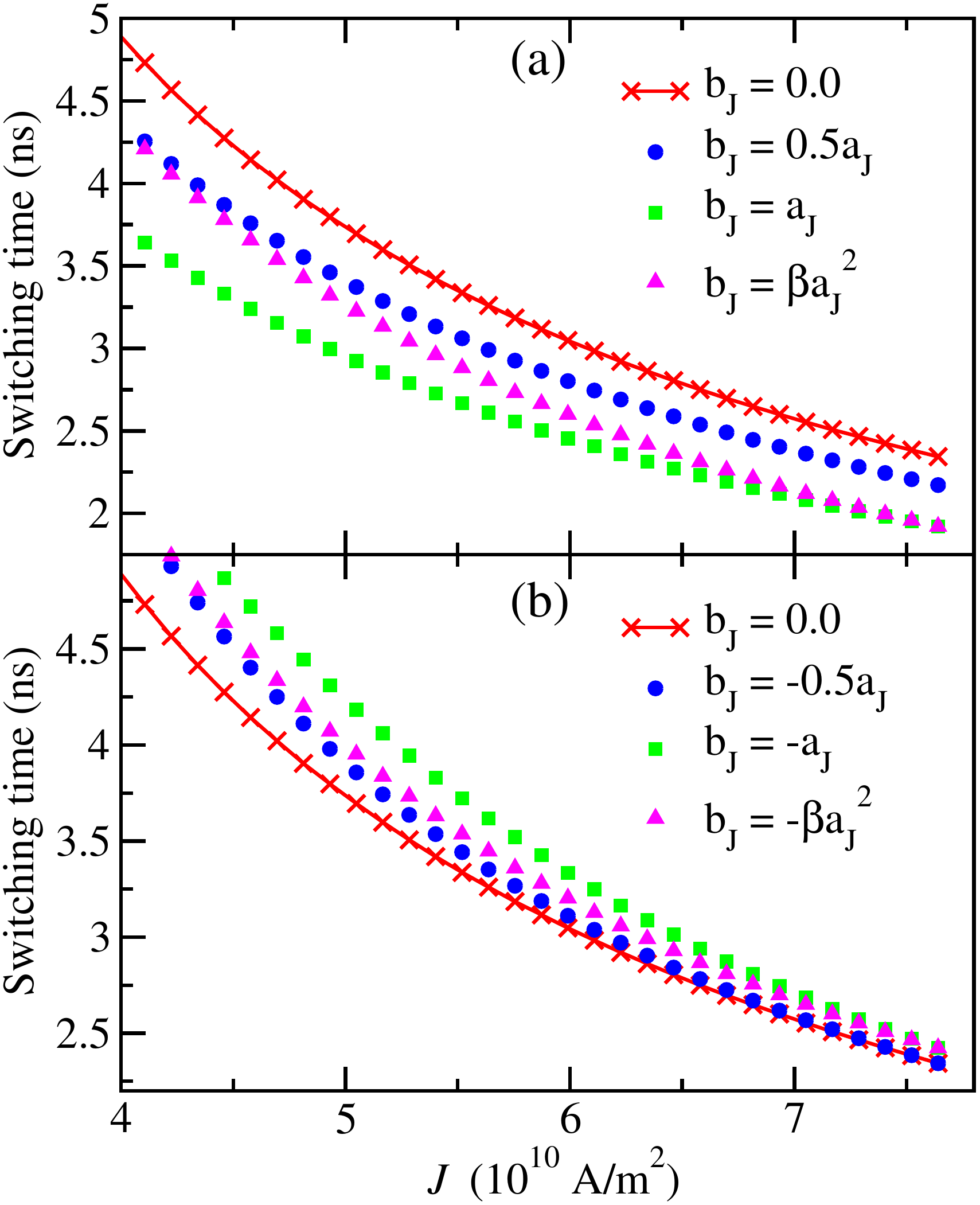}
\caption{\label{fig:figure4}Variation of the switching time 
with the current density $J$ 
when $b_J \propto J$ and $b_J \propto J^2$ 
for (a) positive $\sigma$ and (b) negative $\sigma$.}
\end{figure}

Finally, we investigate how the switching
time changes with current density~\cite{Petit07v98,Sankey08v04,Kubota08v04}  $J$ when (a) $b_J \propto J$
and (b) $b_J \propto J^2$ for positive (negative) 
$\sigma$ in \Fig~\ref{fig:figure4}a (\Fig~\ref{fig:figure4}b)
at $\chi=10^\circ$.
When $b_J \propto J^2$, we work with equivalent
expression $b_J = \beta a_J^2 \ (-\beta a_J^2)$ and choose
$\beta$ value such that $b_J=0.5a_J\ (-0.5a_J)$ 
when $J=4.0\times 10^{10}$~A/m$^2$  and $b_J=1.0a_J\ (-1.0a_J)$
when $J=7.64\times 10^{10}$~A/m$^2$ for $\sigma>0\ (\sigma<0)$.

For positive $\sigma$ both the initial torque as well as the alignment
effect aid in switching; therefore, we see a reduction in the switching
time for all $J$.  Since with increasing magnitude of $\sigma$, both
the effects increase, we see larger reduction in the switching time for
larger $\sigma=1$, \ie, $b_J=a_J$ compared to smaller $\sigma=0.5$.  
Also, when $b_J$ varies quadratically with $J$,
the switching time decreases more rapidly with $J$ compared to the case
when $b_J$ has linear dependence on $J$.  This rapid decrease results
from increasing $\sigma$ with increasing $J$ for the quadratic dependence.

For negative $\sigma$, we see that there is, in fact, an increase in the
switching time in the presence of the field-like term. This is attributable to the
alignment effect (which opposes switching) dominating the initial torque
effect. Also, with increasing $\sigma$, the alignment effect increases
more rapidly compared to the initial torque effect. As a result we see
longer switching time for $\sigma=-1$, \ie, $b_J=-a_J$ compared to that
of $\sigma=-0.5$, \ie, $b_J=-0.5a_J$.  For the quadratic dependence $b_J$ on $J$,
the rate of the switching time increase with $J$ compared to that of
linear dependence because increase in $J$ leads to increased $\sigma$.
Thus we see that the field-like term may improve the switching time if
$b_J$ changes sign the same way as the $a_J$ changes and a quadratic
dependence of $b_J$ with $J$ could lead to a larger reduction as compared
with the linear case.

In summary, we have employed the Landau-Lifshitz-Gilbert equation to
simulate the behavior of a MTJ with a tilted pinned-layer at finite
temperatures, taking the field-like term into account. The field-like
spin-transfer torque $b_J$ is important to the switching dynamics
only if the pinned-layer is tilted. Our simulations illustrate the
simultaneous effect of geometry and the field-like term on switching
distributions at finite temperatures. The present results should allow 
for a more science-based engineering of MTJ  switching performance.

%
\end{document}